# Mass Loss Due to Sputtering and Thermal Processes in Meteoroid Ablation

## L.A. Rogers[1], K. A. Hill & R.L. Hawkes*

Physics Department, Mount Allison University
67 York St., Sackville, NB Canada E4L 1E6

[1]Current address Department of Physics, University of Ottawa, Ottawa ON Canada

*corresponding author: email rhawkes@mta.ca

**Abstract:**

Conventional meteoroid theory assumes that the dominant mode of ablation (which we will refer to as thermal ablation) is by evaporation following intense heating during atmospheric flight. Light production results from excitation of ablated meteoroid atoms following collisions with atmospheric constituents. In this paper we consider the question of whether sputtering may provide an alternative disintegration process of some importance. For meteoroids in the mass range from $10^{-3}$ to $10^{-13}$ kg and covering a meteor velocity range from 11 to 71 km/s, we numerically modeled both thermal ablation and sputtering ablation during atmospheric flight. We considered three meteoroid models believed to be representative of asteroidal (3300 kg m$^{-3}$ mass density), cometary (1000 kg m$^{-3}$) and porous cometary (300 kg m$^{-3}$) meteoroid structures. Atmospheric profiles which considered the molecular compositions at different heights were used in the sputtering calculations. We find that while in many cases (particularly at low velocities and for relatively large meteoroid masses) sputtering contributes only a small amount of mass loss during atmospheric flight, in some cases sputtering is very important. For example, a $10^{-10}$ kg porous meteoroid at 40 km/s will lose nearly 51% of its mass by sputtering, while a $10^{-13}$ kg asteroidal meteoroid at 60 km/s will lose nearly 83% of its mass by sputtering. We argue that sputtering may explain the light production observed at very great heights in some Leonid meteors. We discuss methods to observationally test the predictions of these computations. A search for early gradual tails on meteor light curves prior to the commencement of intense thermal ablation possibly represents the most promising approach. The impact of this work will be most dramatic for very small meteoroids such as those observed with large aperture radars. The heights of ablation and decelerations observed using these systems may provide evidence for the importance of sputtering.

**Introduction:**

Conventional meteor ablation theory (see e.g. McKinley 1961; Ceplecha et al. 1998) assumes that ionization and light production from the atmospheric passage of meteors is a consequence of atomic collisions with ablated meteor atoms. The ablation of meteoroid material is assumed to occur in an intensive manner following the time that the meteor reaches the boiling point. Hence one does not expect luminosity or significant ionization above the height of intensive vaporization, which for most meteors is in the range from 80 to 125 km. The conventional ablation theory has been modified to account for in-flight fragmentation (Jacchia 1955; Verniani 1969; Hawkes & Jones 1975; Ceplecha et al. 1998; Fisher et al. 2000), and to account for differential ablation of meteoroid materials (von Zahn et al.1999), but the assumption remains that ablation is primarily a thermally driven process. Recently Bellot Rubio et al. (2002) have re-analysed precise photographic meteor trails and argue that fragmentation may not be as important as previously presumed.

Öpik (1958) proposed that sputtering, in which individual atoms are released from materials by high speed atomic collision processes, may play some role in meteor ablation. Lebedenits (1970) examined sputtering as the main mass loss mechanism for non-ablating micrometeorites. More recently Brosch et al. (2001) considered simple sputtering and thermal ablation models, and Coulson (2002) and Coulson & Wickramasinghe (2003) have studied sputtering and its importance as a deceleration mechanism to deliver small meteoroids to the Earth with limited in-flight heating. However a comprehensive comparison of thermal processes and sputtering as mass ablation mechanisms has not been published.

In this paper we will provide a detailed analysis of how sputtering based mass loss compares to thermal ablation mass loss for meteoroids in the size ranges which create photographic, image intensified CCD and radar observed meteors. A variety of meteoroid masses, velocities and structural parameters were simulated.

We considered three meteoroid structures in this work. The first (termed asteroidal) corresponded to compact meteoroids reflecting the thermal and mechanical properties of stony meteorites. Asteroidal structure meteoroids had a mass density of 3300 kg/m$^3$. The second (termed cometary) had a bulk mass density of 1000 kg/m$^3$ and is believed to be representative of the majority (in the size range being considered here) of meteoroids which originate from comets. The third structure (termed porous) had a bulk mass density of 300 kg/m$^3$ and is meant to represent cometary material which is very porous due to loss of volatile components prior to atmospheric entry. 300 kg/m$^3$ represents the lower values found in studies of meteoroid density.

**Sputtering:**

Sputtering is considered to be a primary destruction mechanism for interstellar dust (Draine, 1989). It is not part of conventional meteoroid ablation theory, but it has been suggested to be of some importance under certain conditions (Opik, 1958; Lebedenits, 1970; Brosch et al., 2001; Coulson, 2002; Coulson & Wickramasinghe 2003). A summary of the sputtering parameters employed in this work can be found in Table 1.

We are modeling physical sputtering as a possible ablation mechanism for meteoroids traversing the Earth's atmosphere. Physical sputtering is an atomic cascade process whereby bombarding particles (hereafter called projectiles) incident on a target collide with surface atoms, thus dislodging them from the lattice through a transfer of energy. The displaced lattice atoms – as well as the projectile – then undergo collisions with other lattice atoms, dislodging them, and a chain reaction of collisions ensues. Atoms that reach the surface with sufficient energy to overcome the surface potential barrier will escape. The energy needed to overcome the potential barrier is called the surface binding energy ($U_0$). If the velocity component of the atoms normal to the surface corresponds to an energy equal to or greater than the surface binding energy, they will be ejected.

Physical sputtering does not occur for all projectile/target combinations. For a given projectile and target, there exists a minimum projectile kinetic energy needed to induce sputtering. This energy is called the threshold energy ($E_{th}$), and is given by (Bohdansky et al., 1980; Anderson & Bay, 1981)

$$E_{th} = \frac{U_0}{\beta(1-\beta)} \quad \text{for} \quad \frac{M_1}{M_2} \leq 0.3 \quad (1)$$

$$E_{th} = 8U_0 (M_1/M_2)^{1/3} \quad \text{for} \quad M_1/M_2 > 0.3 \quad (2)$$

where $M_1$ is the projectile mass, $M_2$ is the mean molecular mass per atom of the target, and $\beta$, the maximum fractional energy transfer possible in a head-on elastic collision, is given by

$$\beta = \frac{4M_1 M_2}{(M_1 + M_2)^2} \tag{3}$$

For convenience, $U_0$ and $E_{th}$ are usually expressed in electron volts.

For sputtering analysis purposes, it is useful to define a quantity called the sputtering yield, $Y(E,\theta)$, which is the ratio of the mean number of sputtered particles per projectile. The sputtering yield depends on various properties of the target, various properties of the projectile, the incident particle energy, and the incident angle with respect to the surface normal. A comprehensive theoretical relation derived by Tielens et al. (1994) gives the sputtering yield at normal incidence ($\theta = 0$) as

$$Y(E,\theta=0) = \frac{3.56}{U_0(eV)} \frac{M_1}{M_1 + M_2} \frac{Z_1 Z_2}{\sqrt{Z_1^{2/3} + Z_2^{2/3}}} \alpha \frac{R_p}{R} s_n(\gamma) \left[1 - \left(\frac{E_{th}}{E}\right)^{2/3}\right]\left(1 - \frac{E_{th}}{E}\right)^2 \text{ atoms per projectile} \tag{4}$$

valid for $E > E_{th}$, where $Z_1$ is the projectile atomic number and $Z_2$ is the target atomic number. The dimensionless function $\alpha$ depends on the ratio of the atomic masses. For $0.5 < M_2/M_1 < 10$, $\alpha$ is approximated by

$$\alpha = 0.3(M_2/M_1)^{2/3} \tag{5}$$

and remains constant at approximately 0.2 for $M_2/M_1 < 0.5$. The correction factor $R_p/R$ is the ratio of the mean projected range to the mean penetrated path length, and can be approximated by (Bohdansky, 1984)

$$\frac{R_p}{R} = \left(K\frac{M_2}{M_1} + 1\right)^{-1} \tag{6}$$

where *K* will be taken as a free parameter which depends on the target material (see Tielens et al., 1994). The universal function $s_n(\gamma)$, which depends on the detailed form adopted for the screened Coulomb interaction (Tielens et al., 1994), can be approximated by (Matsunami et al., 1980)

$$s_n = \frac{3.441\sqrt{\gamma}\ln(\gamma + 2.718)}{1 + 6.35\sqrt{\gamma} + \gamma(-1.708 + 6.882\sqrt{\gamma})} \tag{7}$$

where $\gamma$ is given by

$$\gamma = \frac{M_2}{M_1 + M_2} \frac{a}{Z_1 Z_2 e^2} E \tag{8}$$

The screening length for the interaction potential between the nuclei (Tielens et al., 1994), *a*, is given by

$$a = \frac{0.885 a_0}{\sqrt{Z_1^{2/3} + Z_2^{2/3}}} \tag{9}$$

where $a_0$ is the Bohr radius. Finally, *E*, which occurs in equations (4) and (8), is simply the kinetic energy of the projectile given by

$$E = \tfrac{1}{2} M_1 v^2 \tag{10}$$

Although $\gamma$ is unitless, *a*, *e* (elementary charge), and *E* must be given in cgs units to remain connsistent with the units in which the sputtering yield equation was derived.

The sputtering yield equation (equation 4) does not take the angle of the projectile (with respect to the surface normal) into account. In the early days of sputtering theory, it was proposed that the sputtering yield exhibits an angular dependence proportional to $(\cos\theta)^{-1}$ (Almen & Bruce, 1961; Molchanov & Tel'kovsky, 1961). More recent work (Jurac et al., 1998) has suggested that for fast moving projectiles, the angular dependence is not proportional to $(\cos\theta)^{-1}$ but is proportional to $(\cos\theta)^{-1.6}$. In this paper, however, only the highest velocities and heaviest projectiles would correspond to energies needing a $(\cos\theta)^{-1.6}$ angular dependence, thus we have used the traditional $(\cos\theta)^{-1}$ dependence. When assuming this angular dependence, limited experimental evidence has suggested that the angle-averaged sputtering yield $\langle Y(E)\rangle_\theta \approx 2Y(E,\theta=0)$ (Draine, 1977; Draine & Salpeter, 1979a), which was successfully used in our model.

The magnitude of the sputtering rate of the meteoroid atoms is given by

$$\frac{dN_{sp}}{dt} = 2Av\left(\frac{m}{\rho_m}\right)^{2/3} \sum n_i Y_i(E) \qquad (11)$$

where $n_i$ and $Y_i$ correspond to the number density (see atmosphere profile section for equations) and sputtering yield of the $i$ different atmospheric molecules bombarding the meteoroid. $A$ is the shape factor of the meteoroid (see McKinley, 1961), while the factor 2 takes the sputtering yield angular dependence into account. The meteoroid rate of change of mass due to physical sputtering can then be written as

$$\left(\frac{dm}{dt}\right)_{sp} = -2M_2 Av\left(\frac{m}{\rho_m}\right)^{2/3} \sum n_i Y_i(E) \qquad (12)$$

While this work is concentrating on meteoroid ablation due to physical (atomic and molecular collision) sputtering, there are also many other types of sputtering mechanisms. These include chemical sputtering, thermal sputtering, electronic sputtering, transmission sputtering, and sputtering by cosmic rays. These sputtering processes have been carefully analyzed for several astrophysical applications (see Barlow, 1978; Draine and Salpeter, 1979), but we did not consider them for sputtering of a meteoroid by the Earth's atmosphere. In addition, it has been suggested that the temperature of a target might influence its sputtering yield, however, that has not been accounted for in the present model.

Guided by the work of Tielens et al. (1994), different sputtering target compositions were adopted for the three meteoroid structures considered in this work. For the asteroidal meteoroids we assumed that the dominant chemical compound was $SiO_2$, while for the cometary and porous meteoroids we assumed that the Tielens carbon (graphite) structure was most appropriate (while recognizing there would also be silicate and metal inclusions). The mean molecular mass per target atom ($M_2$) was weighted so as to encompass the fact that compounds may sputter as atoms and not always as complete molecules. Following Tielens (1994) we adopt $M_2$ values of 20, 12 and 12 for the three structures (asteroidal, cometary, and porous).

**Thermal Ablation:**

As the conventional thermal ablation theory has been well developed by a number of authors (Opik 1958, McKinley 1961, Hawkes and Jones 1975, Nicol et al. 1985, Ceplecha et al. 1998, Fisher et al. 2000), we will describe it in fairly concise terms. The symbol, definition, and values (where applicable) of the parameters used are given in Table 2.

Conservation of linear momentum (see e.g. McKinley 1961) leads to the drag equation which specifies the deceleration of the meteoroid.

$$\frac{dv}{dt} = -\frac{\Gamma A}{m^{1/3} \rho_m^{2/3}} \rho_a v^2 \qquad (13)$$

Here $v$ is the instantaneous speed of the meteoroid and $m$ is the mass of the meteoroid. $\Gamma$ is the drag coefficient which is a dimensionless quantity expressing the efficiency of the collision process in transfer of momentum. It is assumed to have a value not much different from 1. The unitless meteoroid shape factor ($A$) is used to make the equations applicable to meteoroids of any shape; for spherical meteoroids $A$ has a value of 1.21. The density of the meteoroid is given by $\rho_m$ while $\rho_a$ is the atmospheric mass density (which is derived from a model atmosphere – see later in the paper). The gravitational acceleration component can be added to the expression in (13) if desired, but the effect is negligible for the meteoroid velocities considered here.

The atmospheric height ($h$) varies with time according to the relationship

$$\frac{dh}{dt} = -v \cos(z) \qquad (14)$$

where $z$ is the zenith angle of meteoroid entry.

In many previous treatments (e.g. McKinley 1961, Hawkes and Jones 1975, Nicol et al. 1985, Fisher et al. 2000) energy conservation was used to derive an expression for the heating rate of the meteoroid prior to reaching the boiling temperature. After the boiling temperature was reached, it was assumed that there was no further heating and energy conservation was used again to derive the rate of mass loss of the meteoroid. That simple approach neglected a small amount of thermal ablation which occurs prior to the boiling temperature being reached, and is therefore not appropriate for a precise comparison with sputtering which is being used in this paper. Instead we have used the Clausius-Clapeyron equation to model the saturated vapour pressure of the thermally ablating meteoroid. The approach which we follow is essentially that used by Adolfsson et al. (1996) in their study of thermal meteoroid ablation in the Martian atmosphere, which in turn is based on the meteoroid ablation theory as reported by Bronshten (1983). The specific relationships are outlined below.

We have used the meteoroid shape factor, rather than the spherical meteoroid assumption of Adolfsson et al. (1996). In addition, since the meteoroids considered in this work are small, we have assumed isothermal heating which should be valid to first order for these very small meteoroids.

If we apply energy conservation principles to the atmospheric flight of the meteoroid we obtain the following equation.

$$\frac{1}{2}\Lambda\rho_a v^3 = 4\varepsilon\sigma(T^4 - T_a^4) + \frac{cm^{1/3}\rho_m^{2/3}}{A}\frac{dT}{dt} - \frac{L}{A}\left(\frac{\rho_m}{m}\right)^{2/3}\frac{dm}{dt} \qquad (15)$$

In this expression the left hand side represents the energy deposition to the meteoroid from collisions with atmospheric molecules. On the right hand side, the first term represents the net energy loss through thermal radiation, the second term represents the energy which goes into heating of the meteoroid (assumed isothermal in this work as previously mentioned) and the last term represents the energy needed for thermal ablation of meteoroid material. The unitless heat transfer coefficient ($\Lambda$) represents the fraction of the incident kinetic energy which is transferred to the meteoroid, and is usually assumed to have a value near unity. The emissivity of the meteoroid is $\varepsilon$ while $\sigma$ is the Stefan-Boltzmann constant, $T$ is the temperature of the meteoroid while $T_a$ is the effective temperature of the atmosphere in the region of the meteoroid. The meteoroid temperature at the beginning of atmospheric flight was assumed to be 250 K. The specific heat capacity of the meteoroid material is $c$, while the latent heat of fusion plus latent heat of vaporization of the meteoroid material is $L$.

The mass loss rate is determined from the saturated vapor pressure $P_v$ of the thermally ablated meteoroid material according to the following relationship

$$\left(\frac{dm}{dt}\right)_{th} = -4A\left(\frac{m}{\rho_m}\right)^{2/3} P_V(T)\left(\frac{\mu}{2\pi kT}\right)^{1/2} \quad (16)$$

where the Clausius-Clapeyron equation is used to determine the saturated vapour pressure according to the following equation:

$$\log_{10} P_V(T) = C_A - \frac{C_B}{T} \quad (17)$$

In equation (16) $k$ is the Boltzmann constant and $\mu$ is the molecular mass. Unlike sputtering, it was assumed that all products liberated from the meteoroid by thermal ablation were released in complete molecular form. Therefore, the mean molecular masses for thermal ablation were assumed to be 50 (asteroidal $SiO_2$) and 20 for each of the cometary and porous models.

In terms of thermal parameters, we were guided by the work of McKinley (1961), Bronshten (1983), Nicol et al. (1985), Fyfe & Hawkes (1986), Adolfsson et al. (1996) and Fisher et al. (2000), and the references found therein. We use the same thermal parameters for the three different physical meteoroid structures, since the dust component is believed to be approximately the same (see Table 2).

The intensity of meteor radiation ($I$) is assumed to depend on the kinetic energy of the ablated meteor atoms as indicated in equation (18).

$$I = -\frac{1}{2}\tau_I v^2 \frac{dm}{dt} \quad (18)$$

Equations (13) through (18) are solved simultaneously, subject to an atmospheric model (described in detail in the next section).

The luminous efficiency factor, $\tau_I$, represents the fraction of the meteoroid's kinetic energy which has been converted into light energy in the visual range. This factor is one of the most important (since it allows one to relate light produced by the meteor to meteoroid mass), but least conclusively established values in meteor physics. Artificial meteors in the Earth's atmosphere can only be used to measure luminous efficiency at low velocities. Higher speeds can be obtained in the laboratory using electrostatic acceleration but these are restricted to very tiny particles. Some studies have used deceleration and luminosity relationships for bright meteors, but the processes may well be different in these larger meteoroids. In this research we were guided by the relationships given by Jones and Halliday (2001). It should be stressed that the precise value and velocity dependence of luminous efficiency does not have any effect on the main focus of this paper. Figure 1 shows the relationship between the luminous efficiency factor and meteoroid velocity.

**Atmospheric Profile:**

While some of the early meteor ablation studies used simplified constant scale height atmospheres, it is critical to take into account a realistic atmospheric profile. Meteoroid thermal ablation depends on the total atmospheric mass density, while atomic sputtering depends on both the total mass density and the number densities of each atmospheric constituent. We used model atmosphere data to derive equations as a function of altitude for both atmospheric mass density and number densities for individual atmospheric constituents.

All atmospheric data used in creating the atmospheric density profile, which extends from an altitude of 0 to 500 kilometers, was taken from the NASA MSISE-90 model (Hedin 1987, 1991). A representative time of 3 UT and a representative location with a latitude of 45°N and a longitude of 75°W were chosen as input parameters for the model. There would be small changes in the results if a location other than mid northern latitudes was used. An extensive averaging process over seasonal variations and the eleven-year solar cycle was employed to find a set of mean atmospheric densities at each kilometer of altitude. Regressions using several trial functional formats were used to fit the averaged data. There has been evidence that meteor radar rates and solar activity are inversely correlated (Ellyett 1977; Simek & Pecina 2002) with the most likely explanation to be changes in atmospheric density gradient at meteor ablation heights (Ellyett & Kennewell 1980; Lindblad 2003). Therefore while the details of the height of ablation reported here may vary with solar cycle and seasons, our values should represent a reasonable mean behaviour.

To obtain a mean atmospheric mass density, monthly data from the years of 1991 to 2001 were averaged. This range of years spans the eleven-year solar cycle, from solar maximum to the year before the next solar maximum. The regression relationships obtained are presented in Table 3. See Figure 2 for a graphical representation of the mean atmospheric mass density.

As shown in the sputtering equations, the sputtering yield depends on which atmospheric constituent is involved. The atmospheric number densities of $O_2$, $N_2$, He, Ar, O, H, and N were obtained from the MSISE-90 model. Data were obtained for the first day in January, April, July, and October for the years of 1991, 1993, 1995, 1997, 1999, and 2001. These dates were chosen so as to encompass both seasonal and solar cyclic variations in the number densities. The parameters from the resulting equations are given in Table 4 and the number density profiles are plotted in Figure 3.

**Numerical Environment:**

The meteoroid differential equations for sputtering (equations 12-15 and 18 subject to the sputtering rate specified by equations 1-11) and thermal ablation (equations 13 through 18) were solved using a quartic Runge-Kutta technique with a variable step size. Computations were done in Java using double precision mathematics and implemented in a Macintosh programming environment using the Metrowerks CodeWarrior 8.0 Integrated Development Environment. The actual computations were performed in parallel on a number of different processors.

A semi-adaptive step size was applied to the numerical integration in order to achieve a suitable accuracy, while optimizing efficiency. The initial step size ($dt_0$), also assumed as the maximum step size, was calculated using the following formula:

$$dt_0 = \frac{\kappa}{v} \qquad (19)$$

where $v$ is the initial velocity of the meteoroid (m/s) and $\kappa$ is the initial step size constant (units of meters). With a $\kappa$ value of 0.05 very similar results were obtained when $\kappa$ was reduced or increased by a factor of 10. This indicated that sufficient numerical accuracy had been achieved. The chosen value for $\kappa$ ensured that a minimum of twenty steps were calculated for every meter traveled by the meteoroid. As the meteoroid slowed and the step size decreased, however, more steps were calculated per meter.

Once the meteoroid's mass began changing significantly a variable step size was implemented, given by the formula:

$$dt = \frac{Cm}{v\left(\dfrac{dm}{dt}\right)} \qquad (20)$$

where $m$ is the current mass of the meteoroid and $C$ is the adaptive step size constant (units of m/s). After investigating many different values, $C = 0.05$ was chosen because it gave a good balance between efficiency and accuracy (reductions in $C$ were shown to produce similar results within the precision sought).

When a meteoroid first begins to lose mass, $dm/dt$ is initially very small, leading equation (20) above to give excessively large values for $dt$. It was because of this that the value of $dt_0$ was taken as the maximum step size. The adaptive step size, therefore, is not effectively implemented until equation (20) returns a step size $dt < dt_0$. The time interval continues to vary until the meteoroid reaches its maximum light intensity, after which the step size remains constant at its current value until the meteoroid is essentially completely destroyed. This guarantees that the step size remains sufficiently small during the last stages of the meteoroid's ablation while $|dm/dt|$ is decreasing.

The fourth order Runge-Kutta program solves three differential equations related to mass concurrently: $(dm/dt)_{th}$, the rate of mass lost do to thermal ablation, $(dm/dt)_{sp}$, the rate of mass lost due to sputtering, and $dm/dt$ the rate of total mass lost, where

$$\frac{dm}{dt} = \left(\frac{dm}{dt}\right)_{sp} + \left(\frac{dm}{dt}\right)_{th} \qquad (21)$$

At the end of each step, the masses calculated using $(dm/dt)_{th}$ and $(dm/dt)_{sp}$ are used to evaluate the relative contributions of thermal ablation and sputtering to the cumulative and instantaneous mass loss of the meteoroid. Equation (21) was necessary to provide a comprehensive value for the mass of the meteoroid incorporating both sputtering and thermal ablation to be used within each step.

Several criteria were established to determine at which point the program should terminate. Computations were concluded once the meteoroid mass had decreased to 0.00001 of its initial value, signalling the completion of the vast majority of mass loss. A second potential stopping criterion was included so that simulations would terminate when the light intensity decreased to zero. This occurred at meteoroid velocities of approximately 6.2 km/s. A typical run consisted of tens of millions of individual numerical steps. All meteor runs were commenced at an initial altitude of 500 km. The atmospheric densities and compositions reported in the previous section were employed at each step in the process.

**Results:**

In this study we are interested in looking at mass loss mechanisms for meteors which might be detected using photographic, image intensified CCD or meteor radar techniques. Therefore we considered masses of meteoroids from $10^{-3}$ to $10^{-13}$ kg (in increments of a factor of 10). Geocentric velocities of 11.2, 20, 30, 40, 50, 60 and 71 km/s were employed. All results reported here assumed a zenith angle of 45° and a meteoroid shape factor $A = 1.21$ (spherical). Considering the three structures, this meant that a total of 231 different numerical runs were completed for the work reported here (in addition to runs duplicated to verify numerical accuracy under different choices for adaptive step size).

We report in Table 5a the fraction of the mass loss which is due to sputtering (compared to the total mass loss due to both sputtering and thermal processes) for asteroidal composition meteoroids. Sputtering did not occur for 11.2 km/s initial velocity meteoroids since the kinetic energy of the incident atmospheric molecules is less than the threshold energy ($U_0$) for sputtering. As mentioned earlier, our program concluded a run when the meteoroid mass decreased to 0.00001 of the original mass. Not all numerical runs reached this stop criterion. In some of these cases ablation was near complete, while in other cases these were meteoroids for which significant ablation products (or even micrometeorites) remained. When examining the results presented in Table 5a, a relevant quantity to consider is the fraction of the initial meteoroid mass that ablated (due to both sputtering and thermal processes) which is presented in Table 5b. In Table 5b, "1" represents complete ablation while "0" indicates that the meteoroid did not lose any mass within the precision employed. As can be seen, while in many cases (particularly for velocities less than 30 km/s and for relatively large meteoroids) the amount of mass loss due to sputtering is less than 10%, for higher velocities and smaller masses the loss due to sputtering is significant.

Similar results for the cometary model meteoroids are given in Table 6a and Table 6b. For these lower density meteoroids sputtering relative to total ablation is even more important. It should be kept in mind, however, that for the smaller and slower of these meteoroids their structure resulted in significant deceleration and less than total ablation. A typical meteoroid at the limit of a wide field image intensified CCD system (40 km/s velocity, $10^{-8}$ kg meteoroid mass) does completely ablate, with about 18.3% of the total ablation due to sputtering. Clearly detailed ablation models must consider sputtering as part of the process. A cometary meteor near the limit of a modern meteor radar system ($10^{-10}$ kg) typically loses between 25 and 35% of its total mass loss due to sputtering. Ablation by sputtering would be even more important for meteors observed using large aperture radars such as Arecibo. For example, a 50 km/s, $10^{-12}$ kg cometary meteoroid loses 71.3% of its total ablation mass loss through sputtering mechanisms.

Porous meteoroids (reported in Table 7a and Table 7b), as compared to cometary structure meteoroids, in general have only a very small additional contribution from sputtering. In the extreme case of a 71 km/s meteoroid with an initial mass of $10^{-13}$ kg, more than 99% of the total mass loss is due to sputtering. These near hyperbolic orbital velocities in this size range, however, would soon be lost from the solar system by radiation effects.

In Figures 4 and 5 we show the importance of sputtering in the cases of two small porous meteoroids. We ran our ablation programs accounting for only thermal ablation and in the case of both thermal and sputtering ablation. Figure 4 shows the results for a $10^{-8}$ kg porous meteoroid at 40 km/s initial velocity, and would correspond to a meteor which could be detected by conventional meteor radar. The effect of sputtering is clearly observable. Figure 5 is for the extreme case of a very tiny ($10^{-13}$ kg) porous meteoroid entering at 71 km/s, and in this case sputtering totally dominates the ablation process. Meteoroids of this size could only potentially be detected by large aperture radar systems such as Arecibo. As mentioned earlier, however, their solar system lifetimes would be limited.

In Figures 6 and 7 we show, for a subset of the data, the fraction of the entire ablated mass loss which is due to sputtering for cases of asteroidal (3300 kg/m$^3$) and cometary (1000 kg/m$^3$) structures. The importance of sputtering increases with decreasing meteoroid mass, and increases with velocity to moderate velocity values at which point it begins to slowly decline.

**Discussion:**

These results clearly show that sputtering is significant as an ablation mechanism for all but the slowest and heaviest meteors. This significance increases with decreasing meteoroid mass and generally peaks at meteoroid velocities of 30 to 60 km/s. Sputtering is more important for porous or cometary meteoroids than for compact asteroidal material. Clearly any precise model of meteoroid ablation (unless dealing with only very large or very slow meteors) should consider mass loss due to sputtering.

The effects are most important for very small meteoroids, such as those observed using large aperture meteor radar systems (e.g. Janches et al., 2000). The importance of sputtering for very small meteoroids, as suggested recently by Coulson & Wickramasinghe (2003) in their study of meteoroid decelerations and atmospheric heating of small meteoroids, is confirmed in this work. Sputtering may be important in slowing meteoroids so that volatile organic components can be delivered to the Earth's surface with limited heating damage (Coulson & Wickramasinghe 2003).

If the meteoroid is icy the sputtering rate increases very significantly because ice has a much lower surface binding energy than silicate meteoroids (e.g. Tielens et al. 1994). While ice is too volatile to persist for significant periods in meteoroid sized objects in Earth intersecting orbits, very recently ejected cometary material may contain significant ice residues. As such, it may be easiest to find evidence for sputtering in meteoroids recently ejected from comets.

It is possible that sputtering may help to explain the ablation at very high altitudes observed by Fujiwara et al. (1998), Spurny et al. (2000b), and Koten et al. (2001), since as several authors have noted (Elford et al. 1997, Campbell et al. 2000) it is difficult to find a composition which could provide significant ablation at these heights from thermal processes alone. It is important to understand sputtering, since the signature of sputtering at high altitudes could readily be confused with thermal ablation from volatile organics (Steel 1998).

It would be interesting to extend these results to other planetary atmospheres, such as the Martian atmosphere, since sputtering, unlike thermal ablation, depends on atmospheric composition. Furthermore, ice residues in meteoroids could be more important for planets further from the sun because as mentioned earlier, the sputtering yield for ice is very high.

In terms of interpretation of the results presented here, the limitations cited elsewhere in the paper should be kept in mind. We have assumed isothermal meteoroids for the thermal ablation. In the sputtering process, we have assumed no dependence of the sputtering yield on temperature. Recent work by Coulson (2002) has suggested that sputtering could act as a deceleration mechanism for small meteoroids, and this is not accounted for in the present model. For light production, which is not crucial to any of the key results of this paper, we have assumed that thermally ablated and sputtered meteoroid constituents will have the same luminous efficiency factor.

It is important to seek observational evidence for sputtering as a meteoroid ablation mechanism. This will probably be challenging since over most of the ablation profile sputtering and thermal ablation will occur in parallel. However, in the very early part of the meteor trail sputtering occurs before any significant thermal ablation would be expected (since the meteoroid temperature is still relatively low at this point). Therefore a search for low levels of luminosity at high altitudes (in the early part of meteor trails) would be one approach. It is possible that the detections of luminosity from Leonid meteors at great altitudes (see e.g. Fujiwara et al. 1998; Spurny et al. 2000a) is in fact direct evidence for sputtering. Recently Koten et al. (2001) have found several non-Leonid meteors at heights above 150 km.

In Figure 8, we show a plot of luminous energy radiated by the meteor (in Watts) versus height for a $10^{-9}$ kg cometary structure meteoroid. Both thermal and sputtering ablation are taken into account. Searches for evidence of sputtering could look at the low levels of luminosity expected at very great heights. However, many other processes such as fragmentation (see e.g. Fisher et al. 2000) or differential ablation due to the inhomogeneous chemical composition of the meteoroid (see e.g. Borovicka 1999; Zahn et al. 1999) would probably make any such identifications inconclusive.

To demonstrate how difficult this search will be we have examined the detailed light curve for a meteor which is near the limit of current electro-optical detection methods (see e.g. Hawkes 2002 for a review of electro-optical meteor detection). A cometary structure meteor with an initial velocity of 60 km/s and an initial mass of $10^{-8}$ kg (and with a zenith angle of 45°) will have equal contributions from sputtering and from thermal ablation at a height of 116.7 km. At this height the total light production from the meteor would correspond to +8.6 astronomical magnitude. This would be challenging to detect clearly. This meteor later reaches a total brightness of +7.2 magnitude at a height of 108.1 km (but by that time the light production is totally dominated by thermal ablation).

As mentioned briefly earlier, the mean velocities of ablated meteor atoms might be expected to be greater through the sputtering process than for thermal ablation. This might provide a second method to test for sputtering; one could examine the widths of meteor trails at very high spatial resolution. More detailed theoretical work should be undertaken in order to predict sputtered atomic velocities and the resultant widths of meteor trails. It is possible that the width observed in a small number of Leonid meteors is due to this process (e.g. LeBlanc et al. 2000; Spurny et al. 2000b), although we doubt that this process alone can account for the observed widths. It is significant that both the ablation at very great heights and the ablation with significant spatial extent are observed in Leonid meteors shortly after the appearance of the parent comet. It is possible that some of these meteoroids retain enough ice to increase the sputtering yield significantly which could well make sputtering dominate the ablation process.

So far we have concentrated on potential observational proof in the optical detection of meteors. As pointed out in the results section, sputtering is relatively more important for smaller meteoroids. Since high sensitivity meteor radars can observe smaller meteoroids than those observed with image intensified detection systems, they may offer a more promising method to search for sputtering. The main technique may be to search for ionization trails at high altitudes. Unfortunately the rapid expansion of radar trails at these heights, and the resultant interference effects, severely limit the sensitivity of most radars above a certain height. It should be noted here that several authors have argued, based on radar records, that there is very significant meteor input at great heights (e.g. Steel & Elford 1987). While the interpretation was given that there is a significant unobserved high velocity component, a second possible interpretation would be that sputtering is playing a significant role in the meteor ablation process.

The most dramatic sputtering effects would be expected for those meteors which are observed by large aperture radar systems such as Arecibo (Janches et al., 2000). The heights and decelerations observed by these systems may provide a means to search for evidence of sputtering in meteors.

**Acknowledgements:**
The research program of which this forms a part is funded by the Natural Sciences and Engineering Research Council of Canada. We acknowledge in addition computational resources and internal research funding provided by Mount Allison University.

Leblanc, A.G., Murray, I.S., Hawkes, R.L., Worden, P., Cambell, M.D., Brown, P., Jenniskens, P., Correll, R.R., Montague, T., Babcock, D.D., 2000. Evidence for transverse spread in Leonid meteors. Mon. Not. R. astr. Soc. 313, L9-L13.

Lindblad, B. A., 2003. Solar control of meteor radar rates. In: Wilson, A. (Ed.), Solar variability as an input to the Earth's environment. ESA SP-535. 755-759.

Matsunanmi A., Yamamura, Y., Itikawa, Y., Itoh, N., Kazumata, Y., Miyagawa, S., Morita, K., Shimizu, R., 1980. A semiempirical formula for the energy dependence of the sputtering yield. Radiat. Eff. Lett. 57, 15-21.

McKinley, D.W.R., 1961. Meteor Science and Engineering. McGraw-Hill, Toronto.

Molchanov, V.A., Tel'kovsky, V.G., 1961. The dependence of sputtering yield on ion incidence onto the target. Dokl. Akad. Nauk. SSSR. 136, 801-802.

Nicol, E.J., Macfarlane, J., Hawkes, R.L., 1985. Residual mass from atmospheric ablation of small meteoroids. Planet. Space Sci. 33, 315-320.

Öpik, E.J., 1958. Physics of Meteor Flight in the Atmosphere. Interscience, New York.

Simek, M., Pecina, P., 2002. Radar sporadic rates and solar activity. Earth Moon Planets. 88, 115-122.

Spurny, P., Betlem, H., Jobse, K., Koten, P., van't Leven, J., 2000a. New type of radiation of bright Leonid meteors above 130 km. Meteoritics Planet. Sci. 35, 1109-1115.

Spurny, P., Betlem, H., van't Leven, J., Jenniskens, P., 2000b. Atmospheric behavior and extreme beginning heights of the 13 brightest photographic Leonid meteors from the ground-based expedition to China. Meteoritics Planet. Sci. 35, 243-249.

Steel, D., Elford, W.G., 1987. The true height distribution and flux of radar meteors. In: Ceplecha, Z., Pecina, P. (Eds.), Interplanetary Matter (10th ERAM). 193-197.

Steel, D., 1998. The Leonid meteors: compositions and consequences. Astron. Geophys. 39, 24-26.

**Figure Captions**

**Figure 1**

Luminous efficiency factor assumed for radiation in the visible region of the electromagnetic spectrum as a function of velocity (in km/s).

**Figure 2**

Plot of atmospheric mass density versus height for the regression fit shown in Table 1.

**Figure 3**

Plot of atmospheric number densities versus height for various atmospheric constituents according to the regression fit shown in Table 4.

**Figure 4**

This shows the importance of sputtering for a porous structure meteoroid (300 kg/m$^3$ density) having an initial velocity of 40 km/s, a zenith angle of 45° and an initial mass of $10^{-8}$ kg. The results of the ablation of the meteor are shown in the cases where only thermal ablation is considered and when both thermal and sputtering are considered.

**Figure 5**

This shows the importance of sputtering for the extreme case of a porous structure meteoroid (300 kg/m$^3$ density) having an initial velocity of 71km/s, a zenith angle of 45° and an initial mass of $10^{-13}$ kg. The results of the ablation of the meteor are shown in the cases where only thermal ablation is considered and when both thermal and sputtering are considered.

**Figure 6**

Plot of the fraction of the entire ablated mass loss which is due to sputtering as a function of velocity for an asteroidal structure meteoroid (3300 kg/m$^3$ density). Four representative masses are illustrated in the figure. All meteoroids are assumed to have a zenith angle of 45°.

**Figure 7**

Plot of the fraction of the entire ablated mass loss which is due to sputtering as a function of velocity for an cometary structure meteoroid (1000 kg/m$^3$ density). Four representative masses are illustrated in the figure. All meteoroids are assumed to have a zenith angle of 45°.

**Figure 8**

Plot of luminous energy (in W) radiated as a function of height (in m) for a cometary structure meteoroid (1000 kg/m$^3$ density) having an initial velocity of 30 km/s, zenith angle of 45°, and an initial mass of 10$^{-9}$ kg. Thermal ablation begins at 129 km. Observational searches could look for early luminosity on the meteor light curves.

**Table Captions:**

**Table 1**

Symbols and definitions of sputtering parameters in equations (1) – (12) based on work of Tielens et al. (1994).

**Table 2**

Symbols, definitions, and numerical values (where applicable) for thermal ablation parameters in equations (13) – (18).

**Table 3**

Coefficients of the atmospheric mass density equation. The equations have the form $\ln \rho_m = a + bh + ch^2 + dh^3 + e \ln h + f \ln^2 h + g \ln^3 h$, where $h$ is the altitude in meters and $\rho_m$ is the number density in kilograms per cubic meter. The equations were derived using averaged atmospheric data from NASA's MSISE-90 model.

**Table 4**

Coefficients of the atmospheric number density equation. The equations have the form $\ln \rho_m = a + bh + ch^2 + dh^3 + eh^4 + fh^5$, where $h$ is the altitude in meters and $n$ is the number density in particles per cubic meter. The equations were derived using averaged atmospheric data from NASA's MSISE-90 model.

**Table 5a**

Fraction of total mass loss which is due to sputtering for meteoroids of different mass and velocity. An asteroidal composition with mass density of 3300 kg/m$^3$ is assumed. When $v = 11.2$ km/s, sputtering does not occur because (in the reference frame of the meteoroid) the energy of the incident atmospheric molecules is less than the threshold energy ($U_0$) for sputtering.

**Table 5b**

Fraction of meteoroid mass which ablated. An asteroidal composition with mass density of 3300 kg/m$^3$ is assumed. A "1" represents complete ablation while a "0" means that the meteoroid did not lose any mass within the precision employed.

**Table 6a**

Fraction of total mass loss which is due to sputtering for meteoroids of different mass and velocity. A cometary structure with mass density of 1000 kg/m$^3$ is assumed.

**Table 6b**

Fraction of meteoroid mass which ablated. A cometary composition with mass density of 1000 kg/m$^3$ is assumed.

**Table 7a**

Fraction of total mass loss which is due to sputtering for meteoroids of different mass and velocity. A porous cometary structure with mass density of 300 kg/m$^3$ is assumed.

**Table 7b**

Fraction of meteoroid mass which ablated. A porous composition with mass density of 300 kg/m$^3$ is assumed.

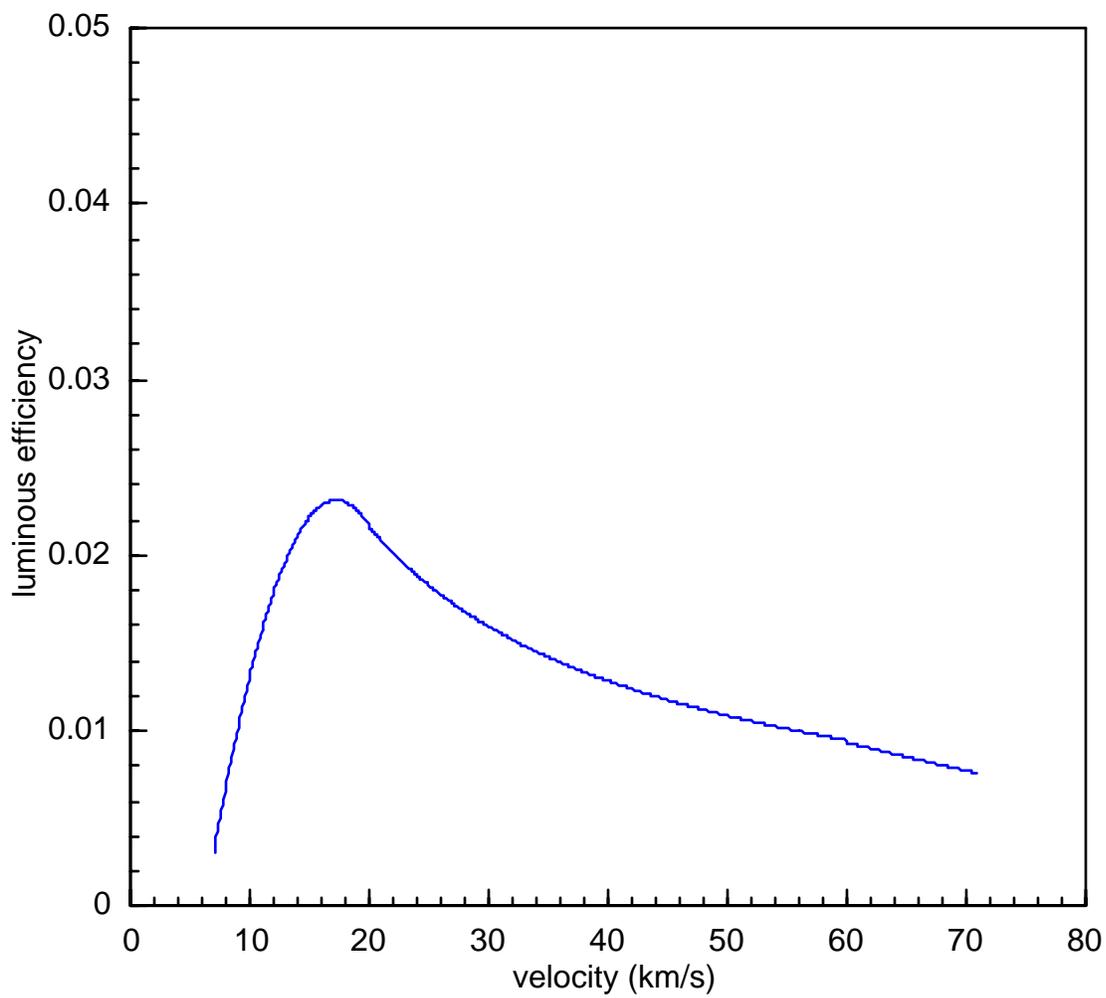

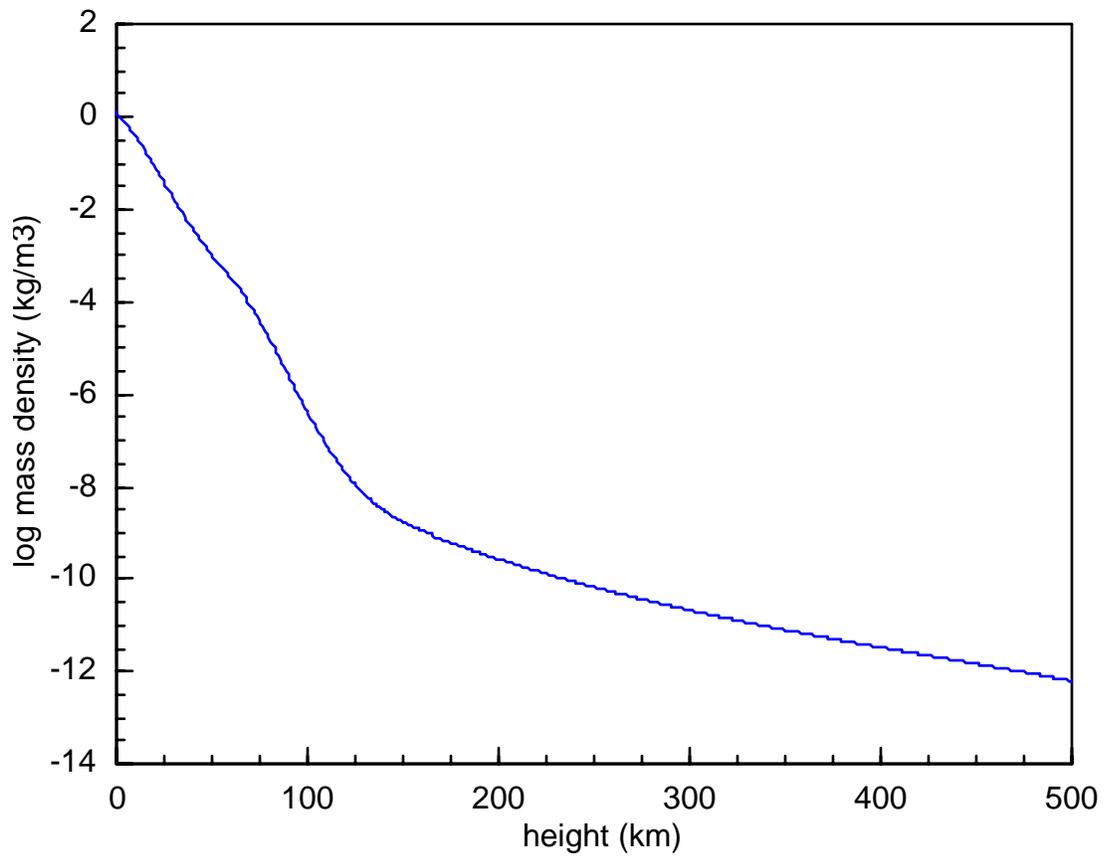

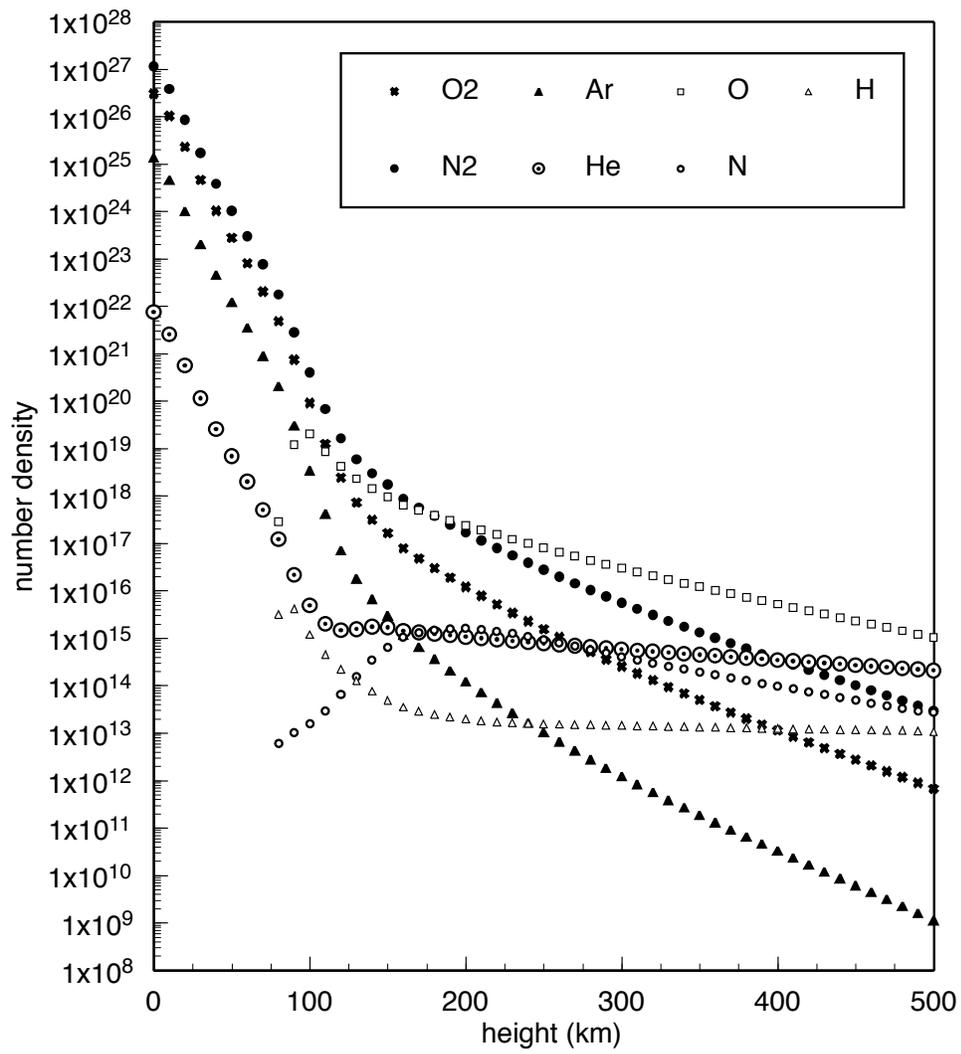

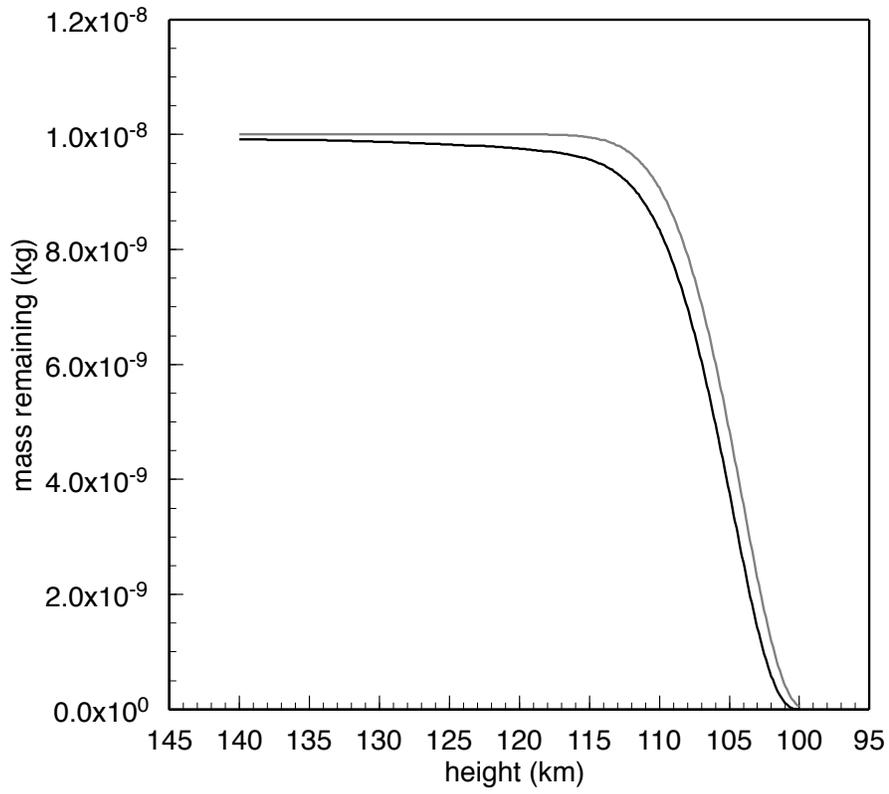

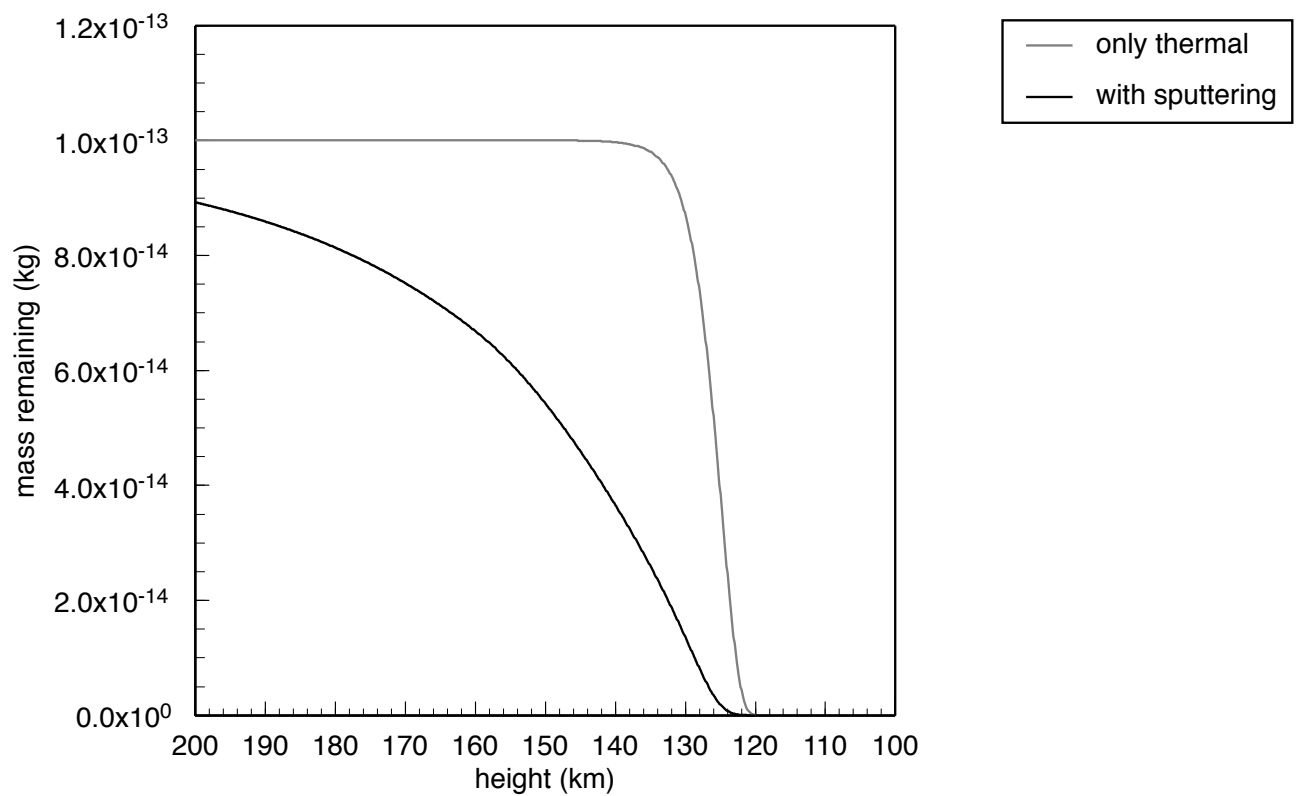

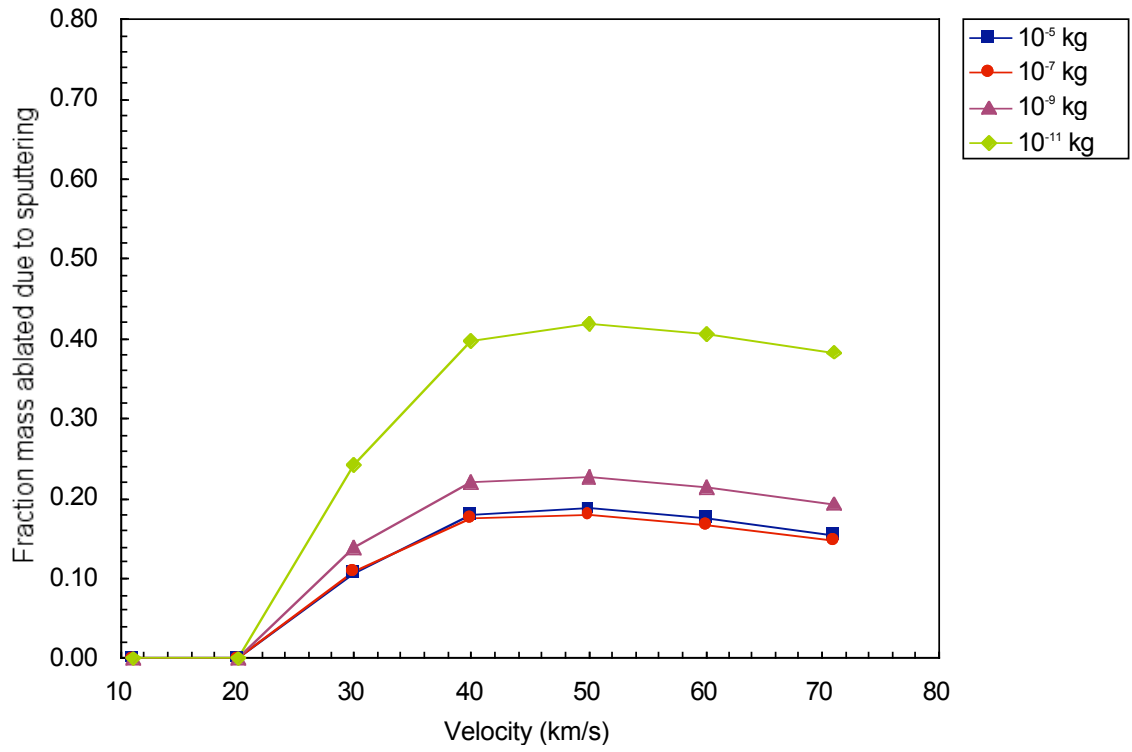

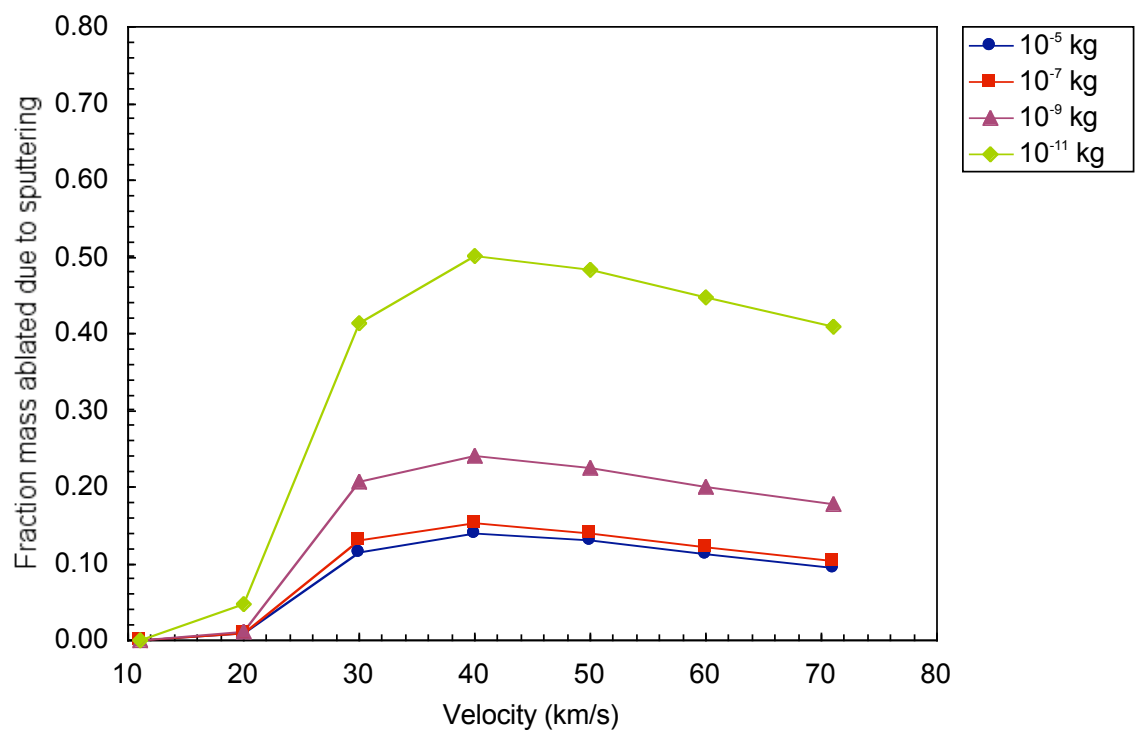

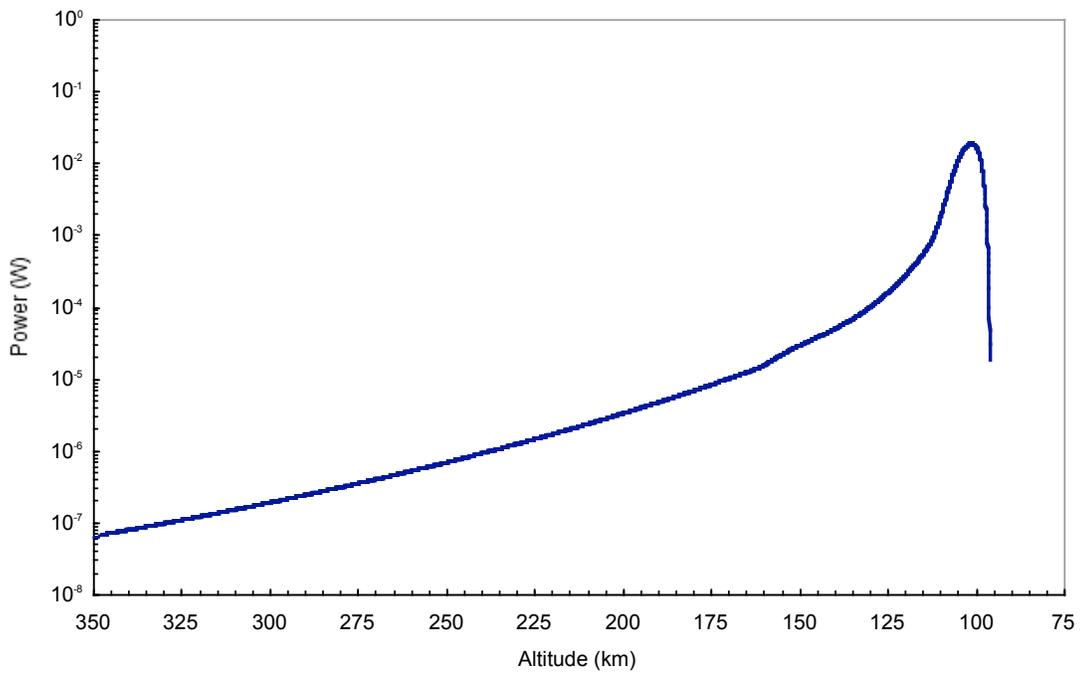

## Table 1: Sputtering Parameters

| Symbol | Definition |
|---|---|
| $U_0$ | Surface Binding Energy |
| $E_{th}$ | Threshold Energy |
| $M_1$ | Projectile Mass |
| $M_2$ | Mean Molecular Mass per Atom of the Target |
| $\beta$ | Maximum Fractional Energy Transfer in Head-On Elastic Collision |
| $Y$ | Sputtering Yield |
| $Z_1$ | Projectile Atomic Number |
| $Z_2$ | Target Atomic Number |
| $\alpha$ | Dimensionless Function of the Mass Ratio |
| $R_p/R$ | Mean Projected Range to Mean Penetrated Path Length |
| $K$ | Free Parameter |
| $s_n$ | Universal Function |
| $\gamma$ | Reduced Energy |
| $a$ | Screening Length |
| $a_o$ | Bohr Radius |
| $e$ | Elementary Charge |
| $E$ | Incident Projectile Energy |
| $N_{sp}$ | Number of Atoms Sputtered from Meteoroid |
| $n$ | Atmospheric Number Density |
| $(dm/dt)_{sp}$ | Meteroid Mass Change Due to Sputtering |

**Table 2: Thermal Ablation Parameters**

| Symbol | Definition | Numerical Value |
|---|---|---|
| $h$ | Meteoroid Height | – |
| $v$ | Meteoroid Velocity | – |
| $T$ | Meteoroid Surface Temperature | – |
| $m$ | Meteoroid Mass | – |
| $(dm/dt)_{th}$ | Meteoroid Mass Change due to Thermal Ablation | – |
| $\rho_a$ | Atmospheric Mass Density | – |
| $\rho_m$ | Meteoroid Density | – |
| $\Gamma$ | Drag Coefficient | 1.0 |
| $A$ | Shape Factor | 1.21 |
| $z$ | Zenith Angle | 45° |
| $\Lambda$ | Heat Transfer Coefficient | 1.0 |
| $\varepsilon$ | Emissivity | 0.9 |
| $\sigma$ | Stefan-Boltzmann Constant | $5.67 \times 10^{-8}$ W/m² K$^{-4}$ |
| $T_a$ | Atmospheric Temperature | 280° |
| $c$ | Specific Heat of Meteoroid | 1200 J/K kg |
| $L$ | Latent Heat of Fusion plus Vaporization | $6.0 \times 10^6$ J/kg |
| $k$ | Boltzmann Constant | $1.381 \times 10^{-23}$ J/K |
| $\mu$ | Mean Molecular Mass of Ablated Vapor | – |
| $P_v$ | Vapor Pressure of Meteoroid | – |
| $C_A$ | Clausius-Clapeyron Coefficient | 10.6 |
| $C_B$ | Clausius-Clapeyron Coefficient | 13 500 K |
| $I$ | Intensity of Visible Meteor Radiation | – |
| $\tau_I$ | Luminous Efficiency Factor | – |
| $\kappa$ | Initial Step Size Constant in Equation (19) | 0.05 m |
| $C$ | Variable Step Size Constant in Equation (20) | 0.05 m/s |

**Table 3: Total Atmospheric Mass Density**

| Altitude Range | Equation Coefficients | | | | | | |
|---|---|---|---|---|---|---|---|
| | a | b | c | d | e | f | g |
| $h < 50000$ | $2.19913 \times 10^{1}$ | $-9.09398 \times 10^{-5}$ | $-2.68411 \times 10^{-9}$ | $3.39694 \times 10^{-14}$ | 0 | 0 | 0 |
| $50000 \leq h < 165000$ | $2.55937 \times 10^{5}$ | $3.21235 \times 10^{-2}$ | $1.91128 \times 10^{-8}$ | 0 | $-7.58756 \times 10^{4}$ | $7.56685 \times 10^{3}$ | $2.54513 \times 10^{2}$ |
| $165000 \leq h \leq 500000$ | $-1.2827 \times 10^{1}$ | $-6.4604 \times 10^{-5}$ | $1.0999 \times 10^{-10}$ | $-8.3612 \times 10^{-17}$ | 0 | 0 | 0 |

## Table 4: Atmospheric Number Densities

| Atmospheric Species | Altitude Range | Equation Coefficients | | | | | |
|---|---|---|---|---|---|---|---|
| | | a | b | c | d | e | f |
| $O_2$ | $h < 73000$ | $6.10057 \times 10^1$ | $-7.12746 \times 10^{-5}$ | $-4.53546 \times 10^{-9}$ | $9.27519 \times 10^{-14}$ | $-5.92157 \times 10^{-19}$ | 0 |
| | $73000 \leq h < 160000$ | $-3.14482 \times 10^1$ | $3.37054 \times 10^{-3}$ | $-4.77364 \times 10^{-8}$ | $2.74203 \times 10^{-13}$ | $-5.64601 \times 10^{-19}$ | 0 |
| | $160000 \leq h \leq 500000$ | $4.90062 \times 10^1$ | $-7.85865 \times 10^{-5}$ | $1.10253 \times 10^{-10}$ | $-8.01809 \times 10^{-17}$ | 0 | 0 |
| $N_2$ | $h < 73000$ | $6.23243 \times 10^1$ | $-7.22646 \times 10^{-5}$ | $-4.46171 \times 10^{-9}$ | $9.09153 \times 10^{-14}$ | $-5.77755 \times 10^{-19}$ | 0 |
| | $73000 \leq h < 160000$ | $-1.47517 \times 10^1$ | $2.85208 \times 10^{-3}$ | $-4.15926 \times 10^{-8}$ | $2.44109 \times 10^{-13}$ | $-5.11749 \times 10^{-19}$ | 0 |
| | $160000 \leq h \leq 500000$ | $5.01497 \times 10^1$ | $-6.84997 \times 10^{-5}$ | $9.49379 \times 10^{-11}$ | $-6.87984 \times 10^{-17}$ | 0 | 0 |
| He | $h < 73000$ | $5.04079 \times 10^1$ | $-7.04124 \times 10^{-5}$ | $-4.62672 \times 10^{-9}$ | $9.63929 \times 10^{-14}$ | $-6.53449 \times 10^{-19}$ | $3.68169 \times 10^{-25}$ |
| | $73000 \leq h < 160000$ | $-2.13316 \times 10^2$ | $1.19401 \times 10^{-2}$ | $-2.13827 \times 10^{-7}$ | $1.82496 \times 10^{-12}$ | $-7.51651 \times 10^{-18}$ | $1.20426 \times 10^{-23}$ |
| | $160000 \leq h \leq 500000$ | $3.64817 \times 10^1$ | $-1.22088 \times 10^{-5}$ | $1.78344 \times 10^{-11}$ | $-1.46951 \times 10^{-17}$ | 0 | 0 |
| Ar | $h < 73000$ | $5.78961 \times 10^1$ | $-7.14397 \times 10^{-5}$ | $-4.52834 \times 10^{-9}$ | $9.27131 \times 10^{-14}$ | $-5.93026 \times 10^{-19}$ | 0 |
| | $73000 \leq h < 160000$ | $-3.63307 \times 10^1$ | $3.42062 \times 10^{-3}$ | $-4.81298 \times 10^{-8}$ | $2.74488 \times 10^{-13}$ | $-5.62124 \times 10^{-19}$ | 0 |
| | $160000 \leq h \leq 500000$ | $4.74112 \times 10^1$ | $-1.01128 \times 10^{-4}$ | $1.55725 \times 10^{-10}$ | $-1.19261 \times 10^{-16}$ | 0 | 0 |
| N | $h < 73000$ | 0 | 0 | 0 | 0 | 0 | 0 |
| | $73000 \leq h < 160000$ | $-1.64974 \times 10^2$ | $8.44859 \times 10^{-3}$ | $-1.44755 \times 10^{-7}$ | $1.21689 \times 10^{-12}$ | $-4.98611 \times 10^{-18}$ | $7.98039 \times 10^{-24}$ |
| | $160000 \leq h \leq 500000$ | $-1.16933$ | $5.71683 \times 10^{-4}$ | $-3.41097 \times 10^{-9}$ | $9.69120 \times 10^{-15}$ | $-1.34938 \times 10^{-20}$ | $7.391632 \times 10^{-27}$ |
| O | $h < 73000$ | 0 | 0 | 0 | 0 | 0 | 0 |
| | $73000 \leq h < 115000$ | $-1.03496 \times 10^2$ | $1.06507 \times 10^{-3}$ | $5.11605 \times 10^{-8}$ | $7.45582 \times 10^{-13}$ | $2.75453 \times 10^{-18}$ | 0 |
| | $115000 \leq h < 160000$ | $7.22170 \times 10^1$ | $-5.03149 \times 10^{-4}$ | $2.84499 \times 10^{-9}$ | $-5.72513 \times 10^{-15}$ | 0 | 0 |
| | $160000 \leq h \leq 500000$ | $4.61301 \times 10^1$ | $-3.90036 \times 10^{-5}$ | $5.03715 \times 10^{-11}$ | $-3.69961 \times 10^{-17}$ | 0 | 0 |
| H | $h < 73000$ | 0 | 0 | 0 | 0 | 0 | 0 |
| | $73000 \leq h < 83000$ | $1.27188 \times 10^3$ | $-5.69682 \times 10^{-2}$ | $8.34563 \times 10^{-7}$ | $-3.94517 \times 10^{-12}$ | 0 | 0 |
| | $83000 \leq h < 160000$ | $-7.63952 \times 10^1$ | $5.67971 \times 10^{-3}$ | $-1.05546 \times 10^{-7}$ | $9.29242 \times 10^{-13}$ | $-3.97120 \times 10^{-18}$ | $6.65029 \times 10^{-24}$ |
| | $160000 \leq h \leq 500000$ | $4.87925 \times 10^1$ | $-2.64976 \times 10^{-4}$ | $1.51842 \times 10^{-9}$ | $-4.30964 \times 10^{-15}$ | $6.02362 \times 10^{-21}$ | $-3.31696 \times 10^{-27}$ |

**Table 5a: Asteroidal Meteoroids (Density = 3300 kg/m³)**
*Fraction of Total Mass Lost Due to Sputtering*

|           | Velocity (km/s) | | | | | | |
|-----------|------|-----------|-------|-------|-------|-------|-------|
| Mass (kg) | 11.2 | 20        | 30    | 40    | 50    | 60    | 71    |
| $10^{-3}$  | 0    | 0.0000998 | 0.103 | 0.183 | 0.199 | 0.189 | 0.171 |
| $10^{-4}$  | 0    | 0.000104  | 0.107 | 0.183 | 0.194 | 0.182 | 0.163 |
| $10^{-5}$  | 0    | 0.000104  | 0.107 | 0.179 | 0.187 | 0.174 | 0.154 |
| $10^{-6}$  | 0    | 0.0000981 | 0.107 | 0.175 | 0.180 | 0.167 | 0.147 |
| $10^{-7}$  | 0    | 0.0000905 | 0.109 | 0.175 | 0.180 | 0.166 | 0.147 |
| $10^{-8}$  | 0    | 0.0000827 | 0.118 | 0.188 | 0.192 | 0.178 | 0.159 |
| $10^{-9}$  | 0    | 0.0000746 | 0.139 | 0.221 | 0.228 | 0.214 | 0.194 |
| $10^{-10}$ | 0    | 0.0000650 | 0.178 | 0.286 | 0.299 | 0.285 | 0.264 |
| $10^{-11}$ | 0    | 0.0000597 | 0.241 | 0.397 | 0.419 | 0.407 | 0.383 |
| $10^{-12}$ | 0    | 0.000138  | 0.332 | 0.567 | 0.603 | 0.594 | 0.568 |
| $10^{-13}$ | 0    | 0.00386   | 0.482 | 0.789 | 0.830 | 0.826 | 0.803 |

**Table 5b: Asteroidal Meteoroids (Density = 3300 kg/m³)**
*Fraction of Mass that Ablated*

|           | Velocity (km/s) | | | | | | |
|-----------|---------|---------|---------|---------|----|----|----|
| Mass (kg) | 11.2    | 20      | 30      | 40      | 50 | 60 | 71 |
| $10^{-3}$  | 0.95050 | 1       | 1       | 1       | 1  | 1  | 1  |
| $10^{-4}$  | 0.94547 | 1       | 1       | 1       | 1  | 1  | 1  |
| $10^{-5}$  | 0.93501 | 1       | 1       | 1       | 1  | 1  | 1  |
| $10^{-6}$  | 0.91223 | 1       | 1       | 1       | 1  | 1  | 1  |
| $10^{-7}$  | 0.85838 | 0.99999 | 1       | 1       | 1  | 1  | 1  |
| $10^{-8}$  | 0.72290 | 0.99992 | 1       | 1       | 1  | 1  | 1  |
| $10^{-9}$  | 0.41590 | 0.99920 | 1       | 1       | 1  | 1  | 1  |
| $10^{-10}$ | 0.06955 | 0.99015 | 1       | 1       | 1  | 1  | 1  |
| $10^{-11}$ | 0.00177 | 0.88019 | 0.99995 | 1       | 1  | 1  | 1  |
| $10^{-12}$ | 0.00001 | 0.30182 | 0.99583 | 1       | 1  | 1  | 1  |
| $10^{-13}$ | 0       | 0.00840 | 0.84272 | 0.99936 | 1  | 1  | 1  |

**Table 6a: Cometary Meteoroids (Density = 1000 kg/m³)**
*Fraction of Mass Lost Due to Sputtering*

| Mass (kg) | Velocity (km/s) | | | | | | |
|---|---|---|---|---|---|---|---|
| | 11.2 | 20 | 30 | 40 | 50 | 60 | 71 |
| $10^{-3}$ | 0 | 0.00870 | 0.114 | 0.145 | 0.140 | 0.124 | 0.106 |
| $10^{-4}$ | 0 | 0.00909 | 0.115 | 0.142 | 0.134 | 0.118 | 0.100 |
| $10^{-5}$ | 0 | 0.00925 | 0.115 | 0.139 | 0.130 | 0.113 | 0.0955 |
| $10^{-6}$ | 0 | 0.00941 | 0.119 | 0.141 | 0.130 | 0.113 | 0.0952 |
| $10^{-7}$ | 0 | 0.00982 | 0.130 | 0.152 | 0.140 | 0.122 | 0.104 |
| $10^{-8}$ | 0 | 0.0104 | 0.157 | 0.183 | 0.168 | 0.148 | 0.128 |
| $10^{-9}$ | 0 | 0.0109 | 0.206 | 0.241 | 0.225 | 0.201 | 0.177 |
| $10^{-10}$ | 0 | 0.0312 | 0.287 | 0.341 | 0.323 | 0.294 | 0.263 |
| $10^{-11}$ | 0 | 0.0464 | 0.414 | 0.502 | 0.483 | 0.448 | 0.409 |
| $10^{-12}$ | 0 | 0.675 | 0.664 | 0.733 | 0.713 | 0.674 | 0.628 |
| $10^{-13}$ | 0 | 0.998 | 0.985 | 0.966 | 0.943 | 0.915 | 0.878 |

**Table 6b: Cometary Meteoroids (Density = 1000 kg/m³)**
*Fraction of Mass that Ablated*

| Mass (kg) | Velocity (km/s) | | | | | | |
|---|---|---|---|---|---|---|---|
| | 11.2 | 20 | 30 | 40 | 50 | 60 | 71 |
| $10^{-3}$ | 0.95050 | 1 | 1 | 1 | 1 | 1 | 1 |
| $10^{-4}$ | 0.94547 | 1 | 1 | 1 | 1 | 1 | 1 |
| $10^{-5}$ | 0.93501 | 1 | 1 | 1 | 1 | 1 | 1 |
| $10^{-6}$ | 0.91223 | 1 | 1 | 1 | 1 | 1 | 1 |
| $10^{-7}$ | 0.85838 | 0.99999 | 1 | 1 | 1 | 1 | 1 |
| $10^{-8}$ | 0.72290 | 0.99992 | 1 | 1 | 1 | 1 | 1 |
| $10^{-9}$ | 0.41590 | 0.99920 | 1 | 1 | 1 | 1 | 1 |
| $10^{-10}$ | 0.06955 | 0.99015 | 1 | 1 | 1 | 1 | 1 |
| $10^{-11}$ | 0.00177 | 0.88019 | 0.99995 | 1 | 1 | 1 | 1 |
| $10^{-12}$ | 0.00001 | 0.30182 | 0.99583 | 1 | 1 | 1 | 1 |
| $10^{-13}$ | 0 | 0.00840 | 0.84272 | 0.99936 | 1 | 1 | 1 |

**Table 7a: Porous Meteoroids (Density = 300 kg/m³)**
*Fraction of Mass Lost Due to Sputtering*

| Mass (kg) | Velocity | | | | | | |
|---|---|---|---|---|---|---|---|
| | 11.2 | 20 | 30 | 40 | 50 | 60 | 71 |
| $10^{-3}$ | 0 | 0.00910 | 0.115 | 0.142 | 0.134 | 0.118 | 0.0998 |
| $10^{-4}$ | 0 | 0.00925 | 0.115 | 0.139 | 0.129 | 0.113 | 0.0954 |
| $10^{-5}$ | 0 | 0.00942 | 0.119 | 0.141 | 0.130 | 0.113 | 0.0954 |
| $10^{-6}$ | 0 | 0.00984 | 0.131 | 0.153 | 0.141 | 0.123 | 0.105 |
| $10^{-7}$ | 0 | 0.0105 | 0.159 | 0.185 | 0.170 | 0.150 | 0.130 |
| $10^{-8}$ | 0 | 0.0109 | 0.208 | 0.244 | 0.228 | 0.204 | 0.180 |
| $10^{-9}$ | 0 | 0.0129 | 0.292 | 0.347 | 0.329 | 0.300 | 0.268 |
| $10^{-10}$ | 0 | 0.0522 | 0.421 | 0.511 | 0.492 | 0.457 | 0.417 |
| $10^{-11}$ | 0 | 0.720 | 0.684 | 0.745 | 0.724 | 0.686 | 0.639 |
| $10^{-12}$ | 0 | 0.999 | 0.988 | 0.972 | 0.951 | 0.924 | 0.888 |
| $10^{-13}$ | 0 | 1.00 | 1.00 | 1.00 | 1.00 | 0.999 | 0.997 |

**Table 7b: Porous Meteoroids (Density = 300 kg/m³)**
*Fraction of Mass that Ablated*

| Mass (kg) | Velocity (km/s) | | | | | | |
|---|---|---|---|---|---|---|---|
| | 11.2 | 20 | 30 | 40 | 50 | 60 | 71 |
| $10^{-3}$ | 0.92858 | 1 | 1 | 1 | 1 | 1 | 1 |
| $10^{-4}$ | 0.89985 | 1 | 1 | 1 | 1 | 1 | 1 |
| $10^{-5}$ | 0.83132 | 0.99998 | 1 | 1 | 1 | 1 | 1 |
| $10^{-6}$ | 0.66182 | 0.99985 | 1 | 1 | 1 | 1 | 1 |
| $10^{-7}$ | 0.31796 | 0.99849 | 1 | 1 | 1 | 1 | 1 |
| $10^{-8}$ | 0.03675 | 0.98134 | 1 | 1 | 1 | 1 | 1 |
| $10^{-9}$ | 0.00077 | 0.79896 | 0.99987 | 1 | 1 | 1 | 1 |
| $10^{-10}$ | 0 | 0.18346 | 0.99068 | 1 | 1 | 1 | 1 |
| $10^{-11}$ | 0 | 0.01254 | 0.78400 | 0.99874 | 1 | 1 | 1 |
| $10^{-12}$ | 0 | 0.00861 | 0.55932 | 0.96329 | 0.99941 | 1 | 1 |
| $10^{-13}$ | 0 | 0.00784 | 0.55573 | 0.95605 | 0.99828 | 0.99996 | 1 |